\documentclass[lettersize,journal]{IEEEtran}
\usepackage[utf8]{inputenc}
\usepackage{cite}
\usepackage{floatrow}
\usepackage{amsmath,amssymb,amsfonts,stfloats}
\usepackage{tabularx}
\usepackage{algorithmic}
\usepackage{graphicx}
\usepackage{textcomp}
\usepackage{floatrow}
\usepackage{xcolor}
\usepackage{booktabs}

\usepackage{multicol}
\usepackage{acronym}
\usepackage{url}
\usepackage{mathtools}
\usepackage[font=small]{caption}
\usepackage[caption=false,font=footnotesize]{subfig}

\def\BibTeX{{\rm B\kern-.05em{\sc i\kern-.025em b}\kern-.08em
    T\kern-.1667em\lower.7ex\hbox{E}\kern-.125emX}}

\begin{document}
 
\title{Target-to-User Association in ISAC Systems With Vehicle-Lodged RIS}

\author{\IEEEauthorblockN{Marouan Mizmizi, Dario Tagliaferri, Damiano Badini, Umberto Spagnolini}}

\maketitle

\begin{abstract}
Target-to-user (T2U) association is a prerequisite to fully exploit the potential of the sensing function in communication-centric integrated sensing and communication (ISAC) systems, e.g., for beam and blockage management. This letter proposes to purposely mount a RIS on the roof of the vehicular user equipment (VUE), which can serve as an intentional back-reflector towards the base station. By controlling the reflection pattern over time, it is possible to transmit information to the sensing system, i.e., back-reflection as bit 1, no back-reflection as bit 0. The VUEs are configured to back-reflect a Hadamard code sequence, which enables T2U association. The numerical results confirm the validity of our proposal. 
\end{abstract}


\section{Introduction}\label{Introduction}

The upcoming 6th Generation (6G) of communication systems is expected to be massively supported by sensing, to enable advanced and novel applications~\cite{Saad2020AVO}. Concerning radio sensing (up to $300$ GHz frequency), the recently emerged paradigm of integrated sensing and communication (ISAC) targets to fuse the two functionalities into a single one with both capabilities~\cite{Zhang2021AnOO}, for full frequency/time/space and hardware resource sharing. Waveform design, obtained via either beampattern optimization~\cite{liu2018toward} or time-frequency resource allocation~\cite{Barneto2021_Optimized_Wave}, is the main research track on ISAC systems and allows trading between communication and sensing performance, enforcing application-specific constraints~\cite{Liu_survey}.

The inherent ISAC trade-off can be tuned to favour either communication or sensing. In the former case, the communication-centric ISAC design is suited to support beam and blockage management (reduction of the beam training time, proactive blockage prediction), resource allocation, etc \cite{Liu_survey}. A vital aspect of a communication-centric ISAC design is the target-to-user (T2U) association procedure, namely recognizing at the ISAC terminal which targets are user equipments (UEs) that are connected over the communication of ISAC. From the perspective of beam management, the importance of T2U association is clear, as the ISAC terminal shall know the number and the position of the UEs to optimally handle the communication (beamforming/beampattern design, resource allocation, etc.). Also, blockage prediction benefits from T2U association as the beam blockage condition applies to UEs only, whereas other targets may only be blockers. Very few works in the available literature address the specific topic of T2U association in ISAC. To the best of the authors' knowledge, the most relevant works in this direction are \cite{Liu2020RadarAssistedPB,Aydogdu2020,FanLiu_TA}. The authors of \cite{Liu2020RadarAssistedPB,Aydogdu2020} focus on sensing-aided vehicular communications and tackle the problem of finding the correct association  among vehicle ID and detected targets using radar and GPS data. The association is carried out by adopting the Kullback-Leibler divergence as a similarity metric to solve a constrained data association problem. Results highlight the beneficial impact of radar data on communication performance. However, environmental factors often affect GPS data, reducing the accuracy and reliability of the proposed work in crowded scenarios. Moreover, exchanging GPS data needs a dedicated and always active low-rate uplink channel. The work in \cite{FanLiu_TA}, instead, overcomes the previous limitation by proposing the usage of ISAC data only for a multi-vehicle tracking algorithm followed by an ID association procedure based again on the Kullback-Leibler divergence between estimated and predicted vehicle states (range, velocity and angle). The authors demonstrate a significant reduction of the uplink ID signalling rate, with a due increase in the achievable rate at each vehicular UE (VUE). Here, there is the issue of discriminating between \textit{collaborative} (VUE) and \textit{non-collaborative} targets, where the latter comprise both targets to be detected and clutter. As demonstrated in \cite{Tebaldini2022}, the environment is rich in static (clutter) and moving targets which are not VUEs and, moreover, vehicles appear as extended targets, thus any association method based on point-target assumption is limited to very low-resolution sensing systems. 

This letter tackles the T2U association issue in ISAC vehicular networks by means of reconfigurable intelligent surfaces (RIS) lodged on vehicles' roofs. RIS are planar or conformal metasurfaces made of sub-wavelength elements whose phase can be dynamically configured to control the reflection properties of the metasurface itself~\cite{DiRenzo2022_RIS_proceedings}. RIS emerged as a promising technology for many applications (see \cite{di2020smart} for a detailed list), but its usage onboard the user is recent \cite{MizmiziConformal2022, Xu2022}. In \cite{Xu2022}, for instance, the authors consider an ISAC system where a base station (BS) illuminates multiple VUEs, each equipped with amplitude- and phase-modulated RIS capable of back reflect a quick response code directly modulating the carrier wave. Differently, we herein propose to use a phase-modulated RIS to encode the VUE ID into the amplitude of the VUEs sensing echo with a proper code, with the specific T2U aim. Using a properly-designed RIS avoids the issues of extended targets while allowing the T2U association from sensing  \textit{without} the fusion with GPS and/or communication ones. 
The numerical results show that a RIS of 10 cm$^2$ in a scenario of $100$ m coverage radius at $70$ GHz operating frequency is sufficient to guarantee a probability of correct detection (PCD) near to $100\%$ in clutter-limited environments. Moreover, the performance of the T2U association, evaluated in terms of the probability of correct association (PCA), shows good robustness to clutter compared to a GPS-based approach. 

The letter is organized as follows: Section \ref{sec:system_model} outlines the system and channel models, Section \ref{sect:EM-Design} details the design of the T2U association procedure based on RIS, Section \ref{sect:results} shows the simulation results and Section \ref{sect:conclusions} concludes the letter.

The notation is as follows: bold upper- and lower-case letters describe matrices and column vectors. Matrix transposition and conjugate transposition are indicated respectively as $\mathbf{A}^{\mathrm{T}}$ and $\mathbf{A}^{\mathrm{H}}$. $\mathbf{I}_n$ is the identity matrix of size $n$. With  $\mathbf{a}\sim\mathcal{CN}(\boldsymbol{\mu},\mathbf{C})$ we denote a multi-variate circularly complex Gaussian random variable $\mathbf{a}$ with mean $\boldsymbol{\mu}$ and covariance $\mathbf{C}$. A similar notation is for log-normal variables $a \sim\mathcal{LN}(\alpha,\beta)$. $\mathbb{E}[\cdot]$ is the expectation operator, while $\mathbb{R}$ and $\mathbb{C}$ stand for the set of real and complex numbers, respectively. $\delta_{n}$ is the Kronecker delta.

\begin{figure}[!b]
    \centering 
    \includegraphics[width=0.7\columnwidth]{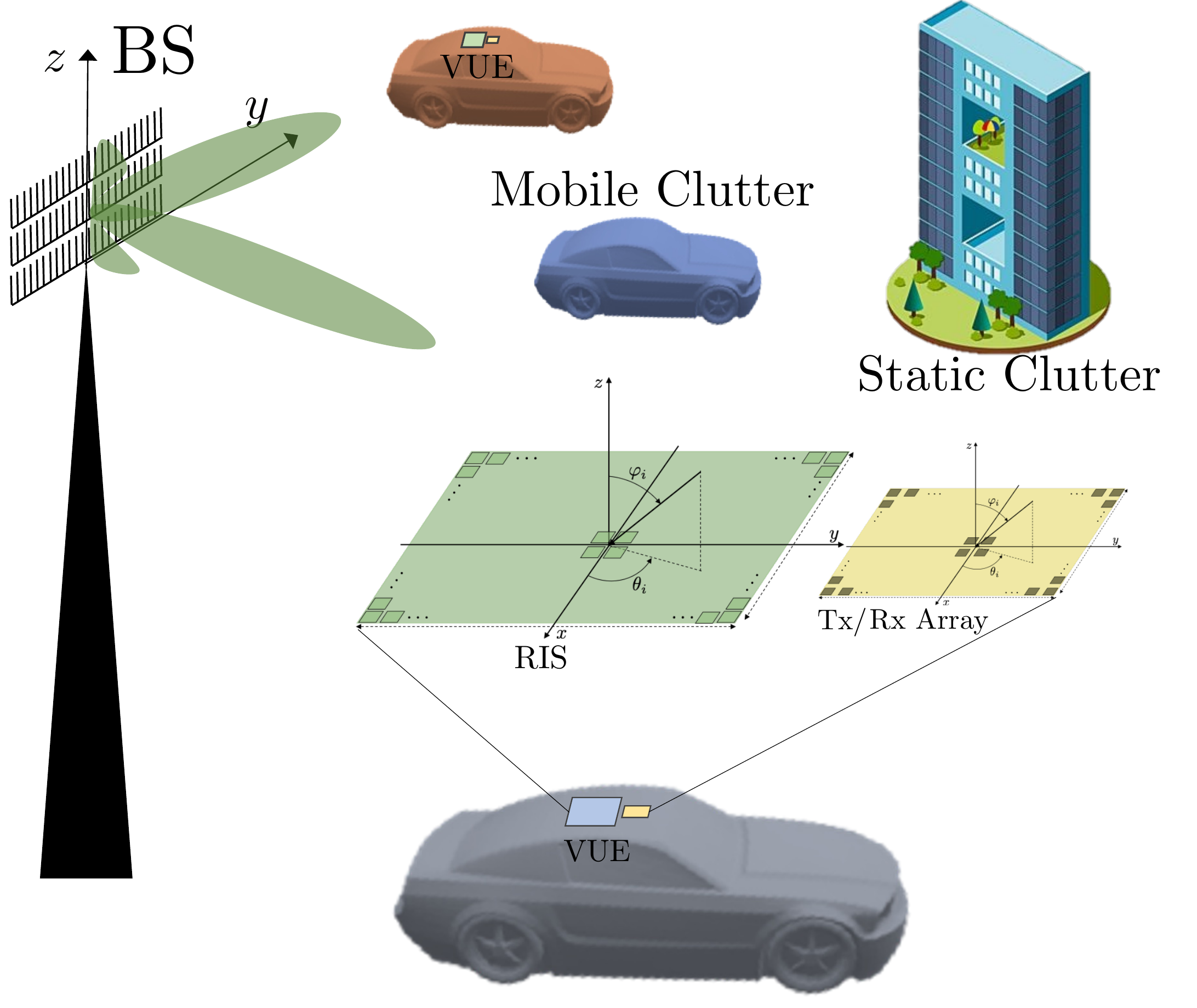}
    \caption{Reference scenario: an ISAC BS is illuminating the environment, composed of VUEs, moving and static targets. The goal is to identify VUEs among all the observed targets}
    \label{fig:intro}
\end{figure}

\section{System and Channel Models}\label{sec:system_model}

Let us consider a downlink ISAC system as depicted in Fig. \ref{fig:intro}. The ISAC node is on a BS, and it is equipped with two antenna arrays of $N$ elements each for sensing the environment while serving $K$ communication UEs, equipped with a single antenna Rx unit (for simplicity) and an $M$-element reconfigurable intelligent surface (RIS). 
The environment is composed of $Q$ targets, comprising $L$ non-collaborative targets, that do not have RIS, and $K$ UEs as collaborative targets. Thus, we have $Q=K+L$. 
The signal emitted by the ISAC-BS is given by
\begin{equation}
    \mathbf{x}(t) = \mathbf{F}\,\mathbf{s}(t)
\end{equation}
where $\mathbf{F}\in\mathbb{C}^{N\times S}$ is the spatial precoding matrix and $\mathbf{s}(t) \in \mathbb{C}^{S\times 1}$ denotes an ISAC waveform with $S$ streams. Signal $\mathbf{x}(t)$ propagates on the communication channel, leading to the Rx signal at the $k$-th UE 
\begin{equation}
    y_k(t) = \mathbf{h}^\mathrm{T}_{k} \mathbf{x}(t-\tau_k)e^{j2\pi\nu_k t} + n_k(t)
\end{equation}
where $\mathbf{h}_{k} \in\mathbb{C}^{N\times 1}$ is the spatial $k$-th communication channel, $\tau_k$ and $\nu_k$ are, respectively, the propagation delay and the Doppler shift from the BS to the $k$-th UE, and $n_k(t)\in\mathcal{CN}(0,\sigma_n^2)$ is the additive white Gaussian noise, uncorrelated across UEs. The delay and Doppler shifts are:
\begin{equation}\label{eq:delay_doppler sen}
        \tau_k = \frac{R_{k}}{c},\,\,\,\,
        \nu_{k} = \frac{f_0}{c} V_{k} 
\end{equation}
where $f_0$ is the carrier frequency, $c$ is the light speed, $R_{k}$ is the range of the UE and $V_{k}$ is its radial component of the velocity.
The $k$-th communication channel is modelled as $\mathbf{h}_{k} = \alpha_k\, \mathbf{a}_N(\boldsymbol{\vartheta}_k)$, where $\alpha_{k}\sim\mathcal{CN}(0,\Omega_k)$ is the channel amplitude, whose power is $\Omega_k \propto (f_0 R_k)^{-2}$ and $\mathbf{a}_N(\boldsymbol{\vartheta}_k)\in\mathbb{C}^{N\times 1}$ is the array response vector of the BS, function of the pointing angles $\boldsymbol{\vartheta}_k$.

The emitted signal $\mathbf{x}(t)$ is also reflected back by the environment, leading to the Rx signal at the ISAC-BS:
\begin{equation}\label{eq:composite_sensing_signal}
\begin{split}
      \mathbf{r}(t)  = &\sum_{k=1}^K  \mathbf{W}^\mathrm{H}\mathbf{H}_{o,k} \boldsymbol{\Phi}_k \mathbf{H}_{i,k} \mathbf{F}\,\mathbf{s}(t-2\tau_k)e^{j4\pi \nu_k t} + \\
    & + \sum_{\ell=1}^L \mathbf{W}^{\mathrm{H}}\mathbf{H}_{\ell}\mathbf{F} \,\mathbf{s}(t-2\tau_\ell)e^{j4\pi \nu_\ell t} + \mathbf{z}(t)  
\end{split}
\end{equation}
where $\mathbf{W}\in\mathbb{C}^{N\times N}$ is the Rx combiner matrix. In \eqref{eq:composite_sensing_signal}, the first term is the collection of the $K$ echoes from the UEs (through the back-reflection from the RIS), the second term is the contribution from the $L$ non-collaborative targets and $\mathbf{z}(t)\sim\mathcal{CN}(\mathbf{0},\mathbf{C}_z(t))$ is
the additive disturbance after the combiner with $\mathbf{C}_z(t)= \sigma^2_z \mathbf{I}_N \delta(t)$. In \eqref{eq:composite_sensing_signal}, $\mathbf{H}_{i,k} \in\mathbb{C}^{M\times N}$ and $\mathbf{H}_{o,k} \in\mathbb{C}^{N\times M}$ are the incident and reflection spatial MIMO channels to/from the $k$-th RIS, modelled as directional: 
\begin{align}
    \mathbf{H}_{i,k} & = \alpha_{i,k}\, \mathbf{a}_M(\boldsymbol{\varphi}_k) \mathbf{a}_N^{\mathrm{H}}(\boldsymbol{\vartheta}_k)\\
    \mathbf{H}_{o,k} & = \alpha_{o,k} \,\mathbf{a}_N(\boldsymbol{\vartheta}_k) \mathbf{a}_M^{\mathrm{H}}(\boldsymbol{\varphi}_k)
\end{align}
where $\alpha_{i,k}$, $\alpha_{o,k}$ are the forward and backward channel amplitudes, whose power is again $\propto (f_0 R_k)^{-2}$, and $\mathbf{a}_M(\boldsymbol{\varphi}_k)\in\mathbb{C}^{M\times 1}$ is the RIS response vector, function of the incident angles $\boldsymbol{\varphi}_k$. The RIS phase configuration matrix is
\begin{equation}\label{eq:refMtx}
    \mathbf{\Phi}_k = \mathrm{diag}\left(\left[e^{j\gamma_{k,0}}, \cdots,  e^{j\gamma_{k,M-1}}\right]\right)\in\mathbb{C}^{M\times M},
\end{equation}
where $\gamma_{k,m}$ is the phase applied to the $m$-th element of the $k$-th RIS. The ISAC-BS arrays have an inter-element spacing of $\lambda_0/2$, while the RIS element spacing is $\lambda_0/4$. As customary, we assume that the sensing channel includes only the two-way LOS path between the BS and each target, and multiple reflections are not considered. For non-collaborative targets, the two-way spatial MIMO channel is 
\begin{equation}\label{eq:channel_model_sensing_TD}
    \mathbf{H}_\ell = \xi_{\ell}\, \mathbf{a}_N(\boldsymbol{\vartheta}_\ell) \mathbf{a}_N^{\mathrm{H}}(\boldsymbol{\vartheta}_\ell)
\end{equation}
with $\xi_{\ell} \sim\mathcal{CN}(0,\Omega_\ell)$ being the scattering amplitude, with $\Omega_\ell\propto(\Gamma_\ell f_0^{-2} R_\ell^{-4})$ (for target's reflectivity $\Gamma_\ell$ and range $R_\ell$). The Doppler shift of the $\ell$-th target, $\nu_\ell$, depends on the parameters as in \eqref{eq:delay_doppler sen}. Herein, the non-collaborative targets are randomly distributed over the coverage area of the ISAC-BS, and their number $L$ is a Poisson random variable with mean value $\rho$ [clutter points/m$^2$]. The log-reflectivity (in dBm$^2$) is modelled with a log-normal distribution as in \cite{DikmenRCS}, thus $\Gamma_\ell\sim\mathcal{LN}(\overline{\Gamma},\sigma^2_\Gamma)$, where parameters $\overline{\Gamma},\sigma^2_\Gamma$ vary to describe different scenarios. The physical reflectivity of the VUEs, when equipped with a perfectly backscattering $M$-element RIS, is \cite{Ellingson2021}
\begin{equation}
    \Gamma_{k}(M) = \frac{4 \pi f_0^2 A^2_{\mathrm{ris}}}{c^2} = \frac{\pi c^2 M^2}{2^6 f_0^2} = \frac{\pi c^2 \|\mathbf{\Phi}_k\|^2_F}{2^6 f_0^2}
\end{equation}
where we approximate the effective area of the RIS as $A_{\mathrm{ris}} = L_{\mathrm{ris}}\times L_\mathrm{ris} = M (\lambda_0/4)^2$ (carrier wavelength $\lambda_0=c/f_0$).

\section{RIS-aided T2U Association}\label{sect:EM-Design}

To enforce the RIS-aided T2U association, necessary to pair communication VUEs with sensed ones, each VUE has a unique ID represented by a binary sequence encoded into the time-varying reflections of the RIS.
The VUE's ID code is implemented by letting the RIS vary its phase configuration as follows: 
\begin{equation}
    \mathbf{\Phi}_k = \begin{dcases}
        \mathbf{\Phi}^{(1)}_k,\quad \text{if the $k$-th RIS is reflecting}\\
        \mathbf{\Phi}^{(0)}_k,\quad \text{if the $k$-th RIS is not reflecting }
        \end{dcases}
\end{equation}
Defining with $T$ the fundamental unit of time of the ISAC system (i.e., either the radar pulse repetition interval or the symbol/slot time of a multicarrier communication waveform), each VUE changes its configuration, switching from $\mathbf{\Phi}^{(0)}_k$ (bit 0) to $\mathbf{\Phi}^{(1)}_k$ (bit 1) or vice-versa, every $T_B = P T$ seconds. The ID code length is herein denoted with $C$, thus the total duration of the ID code is $T_C = C T_B=CPT$. Both $P$ and $C$ parameters rule the T2U association accuracy. Without loss of generality, we employ an orthogonal Hadamard code of length $C\geq K+1$, where we avoid the usage of the all 1 codeword to ease the discrimination of a generic VUE from other non-collaborative targets, whose reflectivity slowly changes over time.
The phase configuration for the RIS in case of full reflection $\mathbf{\Phi}^{(1)}_k$ is based on the knowledge of the incidence angles $\boldsymbol{\varphi}_k=(\theta_k,\psi_k)$, that we herein assume to be perfect. The RIS in this case is configured for back-reflection. For a squared RIS with $\sqrt{M}\times \sqrt{M}$ elements, the phase at the $(u,v)$-th element is
\begin{equation}
    \gamma^{(1)}_{u,v}(\boldsymbol{\varphi}_k) = - \pi \left[ u \, \cos\theta_{k}\sin\psi_{k}
    + v  \,\sin\theta_{k}\sin \psi_{k}\right]
\end{equation}
for $u=0,...,\sqrt{M}-1$, $v=0,...,\sqrt{M}-1$. Differently, the phase for no back-reflection can be achieved by either letting the RIS have a constant phase across the elements (specular reflection) or setting a random and uncorrelated phase at each element. In the former case, the energy back-reflected toward the BS is minimal, but it can generate directional interference. Differently, random spatial phases induce an almost omnidirectional reflection.

\subsection{Target-to-User association accuracy}

To quantify the T2U association accuracy, we can evaluate the probability of correct association (PCA) between the $k$-th UE and the $q$-th target. More specifically, we can define the event in which the $q$-th target is the $k$-th UE, $\mathcal{H}_{q,k}$, and complementary event, $\overline{\mathcal{H}}_{q,k}$.
We define the T2U PCA as:
\begin{equation}\label{eq:PCA}
    \mathrm{P}_\mathrm{ca}^{(k,q)} = \mathrm{P}\left(\widehat{\mathcal{H}}_{k,q}\lvert\mathcal{H}_{k,q},\mathbf{r}(t)\right) \mathrm{P}_\mathrm{cd}^{(q)}
\end{equation}
where $\mathrm{P}\left(\widehat{\mathcal{H}}_{k,q}\lvert\mathcal{H}_{k,q},\mathbf{r}(t)\right)$ is the probability that the ISAC-BS decides to associate the $q$-th target to the $k$-th UE given the received signal $\mathbf{r}(t)$ and that this association is correct (event $\mathcal{H}_{k,q}$), and the second term is the probability of correct detection (PCD) of the $q$-th target. The latter depends on the specific detection method at the Rx and on the disturbance model. For the conventional constant false alarm ratio (CFAR) method, the PCD is upper-bounded by (white noise assumption):
\begin{equation}\label{eq:PCD}
    \mathrm{P}_\mathrm{cd}^{(q)} = \mathcal{QM}_1\left( \sqrt{2\gamma_q},\sqrt{-2 \log \mathrm{P}_\mathrm{fa}} \right)
\end{equation}
where $\mathrm{P}_\mathrm{fa}$ is a target probability of false alarm, $\gamma_q$ is the SNR at the decision variable (i.e., at the input of the ID decoding) and $\mathcal{QM}_1(\alpha,\beta)$ is the QMarcum function of order 1. The PCD in \eqref{eq:PCD} is obtained by optimal thresholding the Rx signal at $\sqrt{-2\sigma^2_z \log(\mathrm{P}_\mathrm{fa})}$, assuming white Gaussian noise with known power as the only source of disturbance.  

The term $\mathrm{P}\left(\widehat{\mathcal{H}}_{k,q}\lvert\mathcal{H}_{k,q},\mathbf{r}(t)\right)$ in \eqref{eq:PCA} is the probability of correct detection of the $k$-th ID code, that can be upper-bounded by
\begin{equation}\label{eq:correction_capability_Hadamard_whitenoise}\mathrm{P}\left(\widehat{\mathcal{H}}_{k,q}\lvert\mathcal{H}_{k,q},\mathbf{r}(t)\right) = \sum_{i=0}^{I} \binom{C}{i}\,p^i (1-p)^{C-i}
\end{equation}
where $p=(1/2)\mathrm{erfc}(\sqrt{\gamma_q/2})$ is the error probability for the single bit and $I= \lfloor \frac{C}{2}-1\rfloor$ is the error correction capability of the Hadamard code. Notice that neither the cross-coupling between VUEs' echoes nor the clutter are taken into account by \eqref{eq:PCD} and \eqref{eq:correction_capability_Hadamard_whitenoise}, that hold for isolated targets and white noise. 
The generic expression of the SNR $\gamma_q$ is in \eqref{eq:SNR_coupledtargets}, where \textit{(i)} the $q$-th target is observed by precoder $\mathbf{f}_q$ and combiner $\mathbf{w}_q$, \textit{(ii)} $T B$ is the time-bandwidth product of the employed ISAC waveform, \textit{(iii)} we average the Rx signal corresponding to the $q$-th target to increase the SNR by $P$ times \textit{(iv)} we neglect the coupling along the range direction.
\begin{figure*}[h!]
    \begin{equation}\label{eq:SNR_coupledtargets}
        \gamma_q = \begin{dcases}
        \frac{\sigma^2_s \lvert \mathbf{w}_q^\mathrm{H}\mathbf{H}_{o,k} \boldsymbol{\Phi}_k \mathbf{H}_{i,k} \mathbf{f}_q\rvert^2 }{ \sum_{j=1,j\neq k}^K  \sigma^2_s \lvert\mathbf{w}_q^\mathrm{H}\mathbf{H}_{o,j} \boldsymbol{\Phi}_j \mathbf{H}_{i,j} \mathbf{f}_q\rvert^2 + \sum_{\ell=1}^L\sigma^2_s \lvert\mathbf{w}_q^\mathrm{H} \mathbf{H}_{\ell} \mathbf{f}_q\rvert^2 + \sigma_z^2/(P T B)} &\text{if $\mathcal{H}_{q,k}$} \\
              \frac{\sigma^2_s \lvert \mathbf{w}_q^\mathrm{H} \mathbf{H}_{\ell} \mathbf{f}_q\rvert^2 }{ \sum_{k=1}^K  \sigma^2_s \lvert\mathbf{w}_q^\mathrm{H}\mathbf{H}_{o,k} \boldsymbol{\Phi}_k \mathbf{H}_{i,k} \mathbf{f}_q\rvert^2 + \sum_{j=1,j\neq q}^L\sigma^2_s \lvert\mathbf{w}_q^\mathrm{H} \mathbf{H}_{j} \mathbf{f}_q\rvert^2 + \sigma_z^2/(P T B)} & \text{if $\overline{\mathcal{H}}_{q,k}$}
        \end{dcases}
    \end{equation}
    \hrulefill
\end{figure*}
Notice that the matched filter gain, in this case, applies only to the noise, as both collaborative and non-collaborative targets are coherent in the observation window of $PT$. The PCD and the PCA in this latter case depend on many parameters, and a general closed-form expression is analytically intractable. 

\section{Results and Discussion}\label{sect:results}

\begin{table}[b!]
\caption{Simulation Parameters}
\centering
\footnotesize
\begin{tabular}{lcc}
\toprule 
\textbf{Parameter} & \textbf{Symbol} & \textbf{Value} \\
\toprule
Frequency & $f_0$ & $70$ GHz\\
Tx power (BS) & $\sigma_s^2$ &  $20$ dBm \\ 
Bandwidth & $B$ & $61$ MHz \\
Cell radius & $R$ &  100 m \\
Noise power level & $\sigma^2_n$ &  $-82$ dBm \\
Number of VUEs & $K$ &  16\\
\bottomrule
\end{tabular}
\label{ParameterSetups}
\end{table} 

In this section, we show the guidelines for RIS design to achieve the T2U association in ISAC-BS. Performance is measured in terms of PCD for the considered scenario. Moreover, through numerical simulations, we prove the benefits of the proposed system in terms of PCA compared to a GPS-based solution. Unless otherwise specified, the simulation parameters in Table \ref{ParameterSetups} are considered.
The precoding matrix $\mathbf{F}$ is designed for both communication and sensing as
\begin{equation}
    \mathbf{F} = [\mathbf{F}_\mathrm{com},\,\mathbf{F}_\mathrm{sens}] 
\end{equation}
where $\mathbf{F}_\mathrm{com}\in\mathbb{C}^{N\times K_\mathrm{com}}$ is a desired precoding optimized for communication with $K_\mathrm{com}\leq K$ VUEs and, similarly, $\mathbf{F}_\mathrm{sens}\in\mathbb{C}^{N\times K}$ is optimized for T2U association, sensing $K$ targets in the environments among the $Q$. Thus, it is $K_\mathrm{com} + K \leq N$. We assume, for simplicity, that the VUEs do not communicate during the T2U association procedure, which enables the optimization of the sensing and communication precoding matrices separately. Nevertheless, the generality of the proposed method is independent of this assumption, and one can use any of the ISAC-optimized precoders discussed in \cite{Zhang2021AnOO}. The combiner is set as $\mathbf{W} = \mathbf{F}_\mathrm{sens}$. 

\subsection{RIS Design}

\begin{figure}[!t]
    \centering
    \includegraphics[width=0.9\columnwidth]{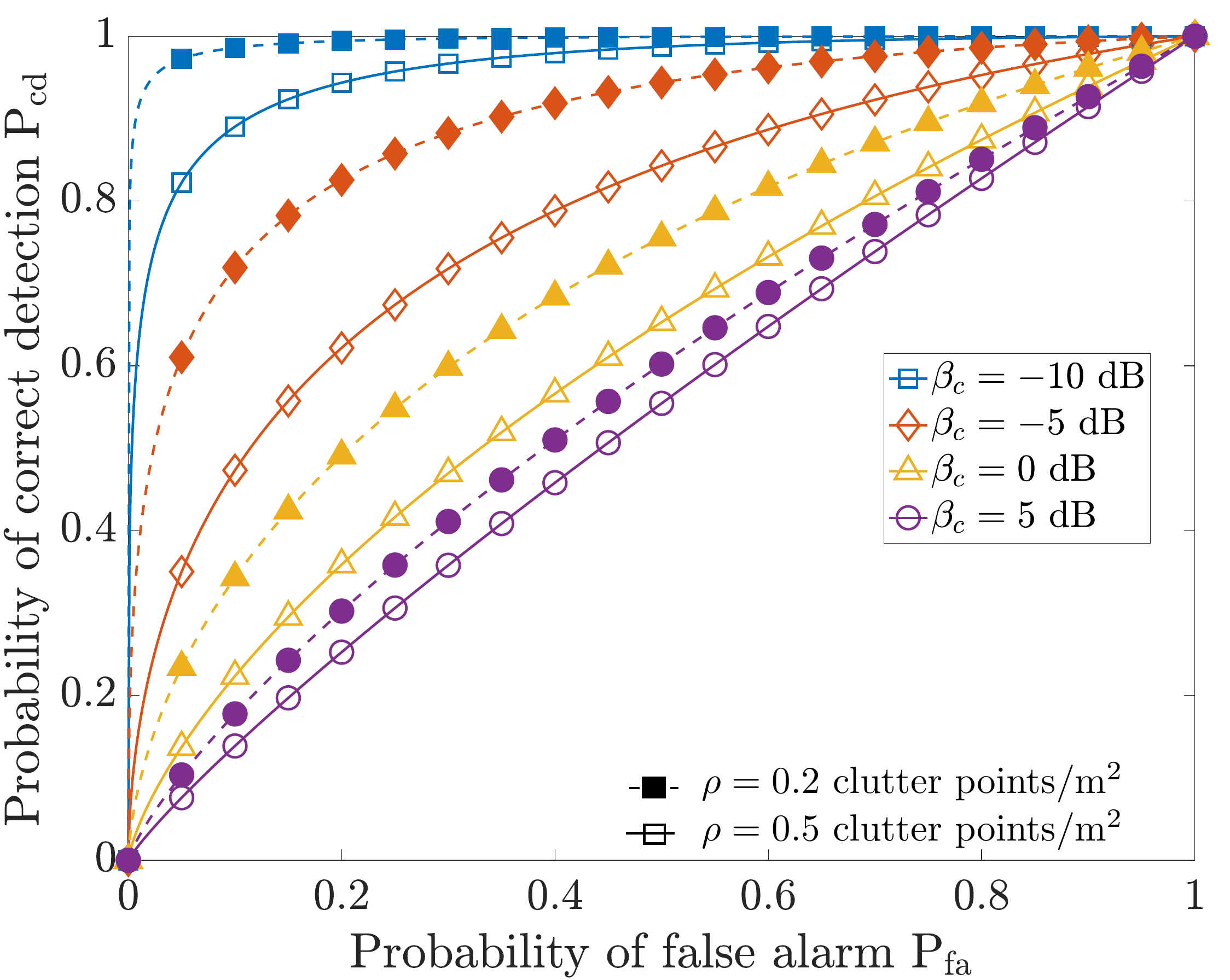}
    \caption{Receiver operating characteristic of the proposed system in presence of clutter with different reflectivity ratio $\beta_c$ and densities $\rho$.}
    \label{fig:ROC}
\end{figure}

Mounting a RIS on top of a vehicle has a major constraint, namely the limited space that restricts its size accordingly. Therefore, it is necessary to understand how large the RIS must be to be visible to the sensing system in the whole scenario. The metric used for measuring the visibility is the PCD upper-bound in \eqref{eq:PCD}, which depends on both sensing SNR in \eqref{eq:SNR_coupledtargets} and the probability of false alarm $P_{\mathrm{fa}}$. Fig. \ref{fig:ROC} depicts the receiver operating characteristic (ROC) curves for two clutter densities ($\rho = 0.2$  dashed lines, and  $\rho = 0.4$ clutter points/m$^2$ solid lines) and considering different reflectivity ratios $\beta_c = \Gamma_\ell / \Gamma_k$ in dB, where the RIS is placed at the cell edge ($R=100$ m). We remark that the simulated clutter densities are general enough to include any source of clutter (e.g., pedestrians, vehicles, etc.) possibly with multiple scattering points (i.e., extended targets). This result suggests that the RIS should have a reflectivity $\Gamma_k$ of at least $10$ dB higher than the clutter reflectivity $\Gamma_\ell$ in the scenario to ensure a $P_{\mathrm{cd}} \geq 90\%$ with a reasonable $P_{\mathrm{fa}} \leq 5\%$. For extremely high clutter density $\rho = 0.4$ clutter points/m$^2$, the requirement becomes stringent.

\begin{figure}[!t]
    \centering
    \includegraphics[width=0.9\columnwidth]{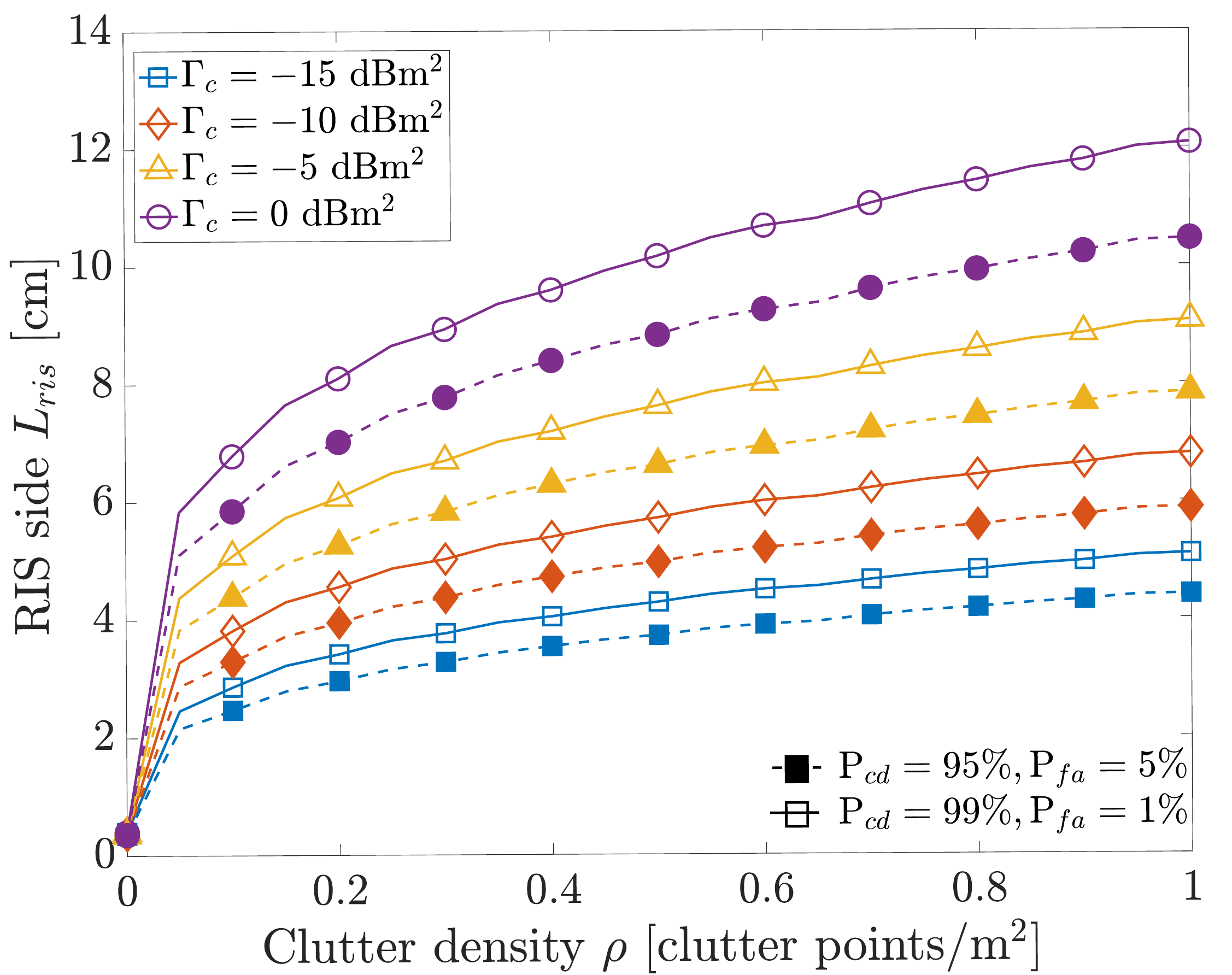}
    \caption{RIS size (side length $L_{\mathrm{ris}}$) varying the clutter density $\rho$, for different clutter reflectivity $\Gamma_c$ and different detection constraints ($\mathrm{P}_{\mathrm{cd}}$ and $\mathrm{P}_{\mathrm{fa}}$).}
    \label{fig:RIS_side}
\end{figure}

Figure \ref{fig:RIS_side} shows the side length $L_{\mathrm{ris}}$ of a squared RIS for varying clutter density $\rho$, required to meet a target $P_{\mathrm{cd}}$ and $P_{\mathrm{fa}}$ during the detection step. The different curves show the trend considering different levels of clutter reflectivity $\Gamma_\ell$. Interestingly, having a RIS with area $A_{\mathrm{ris}} = 10 \times 10$ cm$^2$ ($M\approx 94\times 94$) allows achieving $P_{\mathrm{cd}} \geq 99\%$. Moreover, this result suggests that, under the above conditions, we are far below the size constraint, which for a commercial vehicle is $> 1$ m$^2$.

\subsection{T2U association accuracy}

\begin{figure}[!t]
    \centering
    \includegraphics[width=0.9\columnwidth]{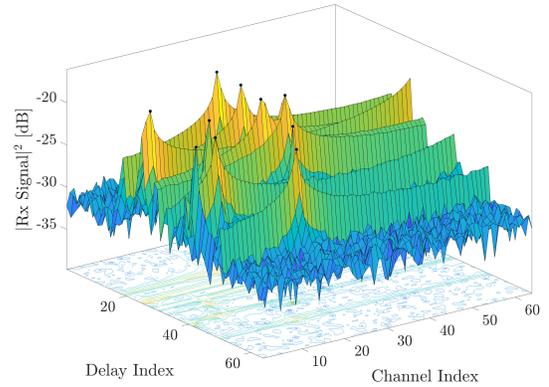}
    \caption{Example of target detection through Fast-Fourier transform and optimal thresholding. Black dots correspond to detected targets.}
    \label{fig:TargetDetection}
\end{figure}

\begin{figure}[!t]
    \centering
    \includegraphics[width=0.9\columnwidth]{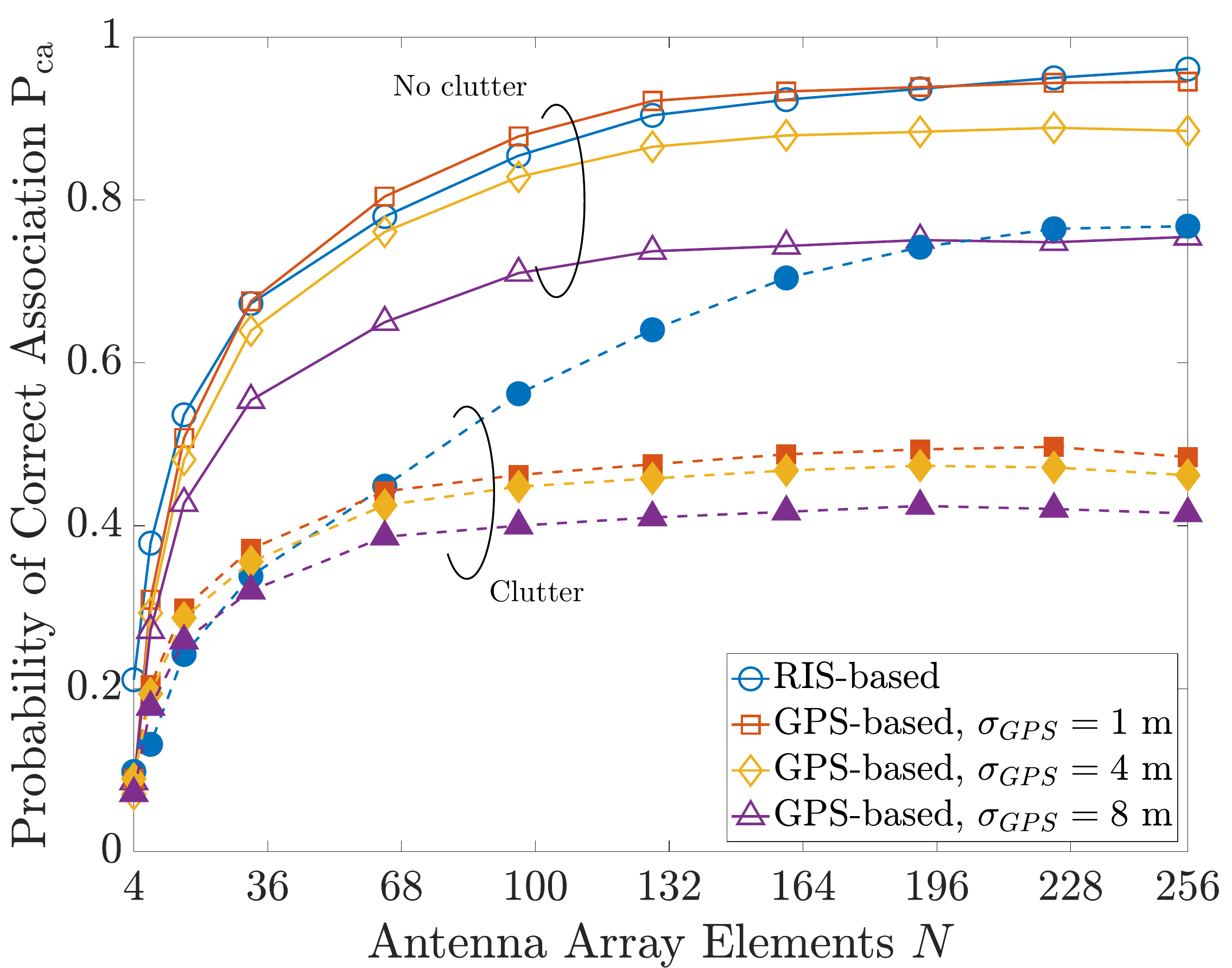}
    \caption{Probability of correct association $\mathrm{P}_\mathrm{ca}$ varying the spatial resolution of the ISAC-BS (number of array elements), with and without clutter. The RIS-based PCA is compared with GPS based approach for different values of GPS positioning accuracy.}
    \label{fig:Pca}
\end{figure}

The benefits of the proposed RIS-assisted T2U procedure are assessed in Fig. \ref{fig:Pca}, showing the PCA varying the number of array elements @ ISAC-BS $N$. We report both the case with clutter (static and mobile) with spatial density $\rho=0.1$ clutter points/m$^2$ and reflectivity $\Gamma_\ell=8$ dBm$^2$ (RIS has $\Gamma_k=18$ dBm$^2$, $\forall k$)
and without. 
We compare the performance of the proposed method with a GPS-aided solution, where the T2U association is obtained with the nearest neighbour algorithm applied to the set of detected targets and GPS measurements by the VUEs. Here, we assume the GPS 3D positions as Gaussian random variables with the VUE exact position as the mean value and $\sigma_{GPS} = [1, 4, 8]$ m as the standard deviation, along each coordinate. Notice that GPS-based T2U techniques require an available BS-VUE control link.  
An example of target detection in the considered setting is shown in Fig. \ref{fig:TargetDetection}. For both systems, we observe an increase in performance as the sensing resolution increases. This is expected for GPS-based approaches, where increasing the sensing spatial resolution allows for better distinguishing VUEs from clutter points within the GPS uncertainty region, easing the association through the minimum distance. For the RIS-based approach, instead, the PCA performance is only ruled by the sensing resolution, as the RIS makes VUEs behave like point targets, enhancing their visibility for sensing.
In absence of clutter, the performance of the RIS-based method matches with the high-accuracy GPS-based one. However, the performance of the GPS-based approach is environment-dependent and drops significantly when the clutter is present, showing low robustness in more realistic scenarios.

\section{Conclusion}\label{sect:conclusions}
 
In the context of communication-centric ISAC systems, this letter tackles the problem of T2U association, which is a prerequisite to fully exploit the potential of the sensing function as a support to the communication system, e.g., for beam or blockage management. The proposed system consists of mounting RIS on the rooftop of VUEs, which can serve as intentional back-reflectors towards the BS, hence enhancing their visibility with sensing. By controlling the reflection pattern over time, it is possible to transmit information to the sensing system. In the proposed solution, the VUEs are configured to reflect a Hadamard code sequence. Our findings show that a RIS of 10 cm$^2$ ensures good performance in terms of PCD and PCA, with good robustness to clutter, and remarkable advantages with respect to positioning-aided solutions.

\section*{Acknowledgment}
The work has been framed within the Huawei-Politecnico di Milano Joint Research Lab. 

\bibliographystyle{IEEEtran}
\bibliography{Bibliography}

\end{document}